\documentclass{elsart}
\usepackage{epsfig}
\usepackage{amssymb}
\usepackage{amsfonts}
\usepackage{here}
\begin{document}
\begin{frontmatter}
\title{Near threshold production of the $\eta$ meson via the quasi-free $pn\to pn\eta$ reaction}
\author[cracow,julich1]{P.~Moskal\corauthref{corr}},
\ead{p.moskal@fz-juelich.de}
\author[cracow]{R.~Czy\.zykiewicz},
\author[munster]{H.-H.~Adam},
\author[insbruck]{S.~D.~Bass},
\author[bronowic]{A.~Budzanowski},
\author[cracow,julich1]{E.~Czerwi\'nski},
\author[cracow]{D.~Gil},
\author[julich1]{D.~Grzonka},
\author[cracow,julich1]{M.~Janusz},
\author[cracow]{L.~Jarczyk},
\author[uppsala]{T.~Johansson},
\author[cracow]{B.~Kamys},
\author[munster]{A.~Khoukaz},
\author[julich1]{K.~Kilian},
\author[cracow,julich1]{P.~Klaja},
\author[cracow,julich1]{J.~Majewski},
\author[julich1]{W.~Oelert},
\author[cracow]{C.~Piskor-Ignatowicz},
\author[cracow,julich1]{J.~Przerwa},
\author[cracow]{B.~Rejdych},
\author[julich1]{J.~Ritman},
\author[katowice]{T.~Ro\.zek},
\author[julich1]{T.~Sefzick},
\author[katowice]{M.~Siemaszko},
\author[cracow]{M.~Silarski},
\author[cracow]{J.~Smyrski},
\author[munster]{A.~T\"aschner},
\author[julich1]{M.~Wolke},
\author[julich1]{P.~W\"ustner},
\author[cracow]{M.~J.~Zieli\'nski},
\author[katowice]{W.~Zipper},
\author[cracow]{J.~Zdebik}
\address[cracow]{Institute of Physics, Jagiellonian University, Pl-30-059 Cracow, Poland}
\address[julich1]{IKP \& ZEL, Forschungszentrum J\"ulich, D-52425 J\"ulich, Germany}
\address[munster]{IKP, Westf\"alische Wilhelms-Universit\"at, D-48149 M\"unster, Germany}
\address[insbruck]{Institute for Theoretical Physics, University of Innsbruck, Austria}
\address[bronowic]{Institute of Nuclear Physics, Pl-31-342 Cracow, Poland}
\address[uppsala]{Department of Physics and Astronomy, Uppsala University, Sweden}
\address[katowice]{Institute of Physics, University of Silesia, PL-40-007 Katowice, Poland}
\corauth[corr]{Corresponding author. Correspondence address: Institute of Physics,
Jagiellonian University, ul. Reymonta 4, Pl-30-059 Cracow, Poland. 
}
\begin{abstract}
Total cross sections for the quasi-free $pn\to pn\eta$
reaction in the range from the kinematical threshold 
up to 20~MeV excess energy have been determined. At threshold they exceed
corresponding cross sections for the $pp\to pp\eta$ reaction  by a factor of about
three in contrast to the  factor of six established  
for higher excess energies. To large extent, the observed decrease 
of the ratio $\sigma(pn \to pn\eta)/\sigma(pp\to pp\eta)$ towards threshold 
may be assigned to the different
energy dependence of the proton-proton and proton-neutron
final state interactions.

The experiment has been conducted using a proton beam of the cooler synchrotron COSY
and a cluster jet deuteron target.  The proton-neutron reactions were tagged by the 
spectator proton whose momentum was measured for each event.  Protons and neutron
outgoing from the $pn\to pn\eta$ reaction have been registered by means 
of the COSY-11 
facility, 
an apparatus dedicated for threshold meson production.
\vspace{1pc}
\end{abstract}
\end{frontmatter}

\section{Introduction}

In recent years meson production has been
extensively studied in the context of understanding of the
strong interaction in the non--perturbative energy 
domain of the quantum chromodynamics,
where it is not obvious whether hadronic or quark-gluon degrees
of freedom are more appropriate for the description
of the dynamics and interaction of hadrons.
 For the $\eta$ meson 
 precise data~\cite{smyrski,chiavassa,calen1,calen2,c11PhysRev,hibou,hab}
  of the total
  cross section of the 
  $pp\to pp\eta$ reaction allowed one to conclude that the reaction proceeds
  predominantly through the excitation of one of the protons to the $S_{11}(1535)$ state which
  subsequently de-excites via emission of the $\eta$ meson. 
  The crucial observations
  were a large value of the absolute cross section (about forty times larger than for the
  $\eta^\prime$ meson~\cite{eta-prime}) 
  and  isotropic angular distributions~\cite{c11PhysRev,hab,tof} of the 
  $\eta$ meson emission in the reaction center-of-mass system.

  However, due to the negligible variation of the production
  amplitude in the range of few tens of MeV the full information
  available from the excitation function is reduced to a single number~\cite{moalem}.
Consequently, measurements of one reaction channel
are not sufficient 
to establish  contributions from different
production currents
(e.g. mesonic, nucleonic, resonance or gluonic~\cite{colin,nakayama,bass1,bass2}).
For this purpose, 
an exploration of isospin and spin degrees of freedom
is mandatory. 
The first measurement of the dependence of the $\eta$
production on the isospin
of the interacting nucleons was conducted
by the WASA/PROMICE collaboration~\cite{calenpn}
in the excess energy range from 16~MeV to 109~MeV.
On the hadronic level, the observed strong isospin dependence 
indicated a dominant contribution to the production
process from the  isovector meson exchange.
Further
comparison of predictions 
involving exchanges of various mesons~\cite{colin,nakayama}  with recent
results on the analysing power~\cite{rafal}
signified 
the dominance of the $\pi$ meson exchange
in the production process
(though with a rather low statistical significance)~\cite{rafal}.
This observation is in line  with predictions of Nakayama et al.~\cite{nakayama}
and with calculations of Shyam~\cite{shyam}.
Yet, 
it seems to be contra-intuitive due to the very large momentum
transfer between the interacting nucleons needed to create the $\eta$ meson
near threshold which should rather favor exchanges of heavier mesons 
(e.g. $\rho$) as  anticipated by authors of references~\cite{colin,colindelta}.
Moreover, with the high momentum transfer a direct role of quarks and gluons  
in the production process is also conceivable.
According to the Heisenberg uncertainty relation
the
distance probed by the $NN \rightarrow
NN\eta$ reactions at threshold is expected to be about 0.26~fm.
Geometrical considerations presented by Maltman and Isgur~\cite{maltmanisgur}
indicate
that at distances smaller than $2\,\mbox{fm}$ the inter-nucleon potential
should begin to be free of meson exchange effects and may be dominated by 
residual colour forces~\cite{maltmanisgur}.
In particular,
since the singlet components of the $\eta$ and $\eta'$ mesons
couple to glue,
it is natural to consider
a mechanism 
where glue is excited in the
``short distance'' ($\sim 0.2$~fm)
interaction region of a proton-nucleon collision and then evolves to
become an $\eta$ or $\eta'$ in the final state.
Such gluonic induced production, proposed by Bass~\cite{bass1},
is extra to the
contributions
associated with
meson exchange models, 
and appears as a contact term
in the axial U(1) extended chiral Lagrangian for low-energy QCD.

In order to verify the different production mechanisms it is of crucial importance
to provide an empirical base with  spin and isospin dependence of the cross
sections  of the near threshold  meson production in
the collision of nucleons as well as in the photo-production processes. 
Therefore, 
such investigations are conducted at several experimental
facilities~\cite{tofpn1,tofpn2,ankepn1,c11pn,jaegle}.
In this article we report on the measurement
which enables us for the first time to determine  cross sections
for the $pn\to pn\eta$ reaction  close to the kinematical threshold
where a contribution from only a single partial wave is expected.

\section{Experimental method}
\begin{figure}[h]
\begin{center}
  \includegraphics[width=7cm]{moskal_pneta_fig1a.eps}
\begin{minipage}{0.54\textwidth}
\vspace{-2cm}
\includegraphics[width=0.45\textwidth]{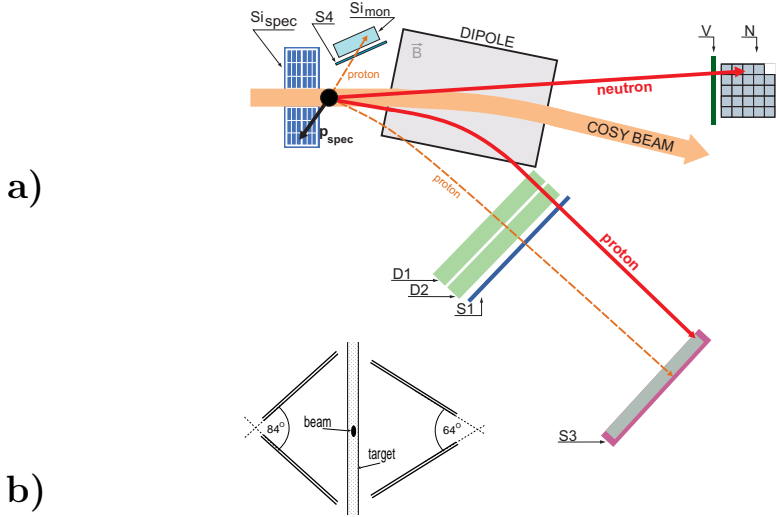}
\end{minipage}
\end{center}
\begin{picture}(15.0,7.5)
\put(6.5,135.0){{\large {\bf a)}}}
      \put(6.0,20.0){{\large {\bf b)}}}
\end{picture}
\caption{ (a) Schematic view of the COSY-11 detection setup~\cite{cosy11}.
D1 and D2 represent drift chambers.
S1, S3, S4 denote  scintillator counters. 
N stands for the neutral particle detector and V for the veto counter. 
Si$_{mon}$ and Si$_{spec}$ 
are silicon strip detectors used for the detection 
of the elastically scattered protons and spectator protons, respectively. 
Superimposed solid lines indicate correspondingly spectator proton, neutron and forward scattered proton
from the quasi-free $pd \to p_{sp} p n \eta$ reaction  and the dashed lines show 
an example of the quasi-free elastically scattered protons used 
for the monitoring of the luminosity.
The size of detectors and their relative distances are not to scale.
(b) Projection of the arrangement of the spectator detector 
as seen along the beam axis. The beam is moving in the direction
perpendicular to the plane of the projection, while the target deuterons are moving 
from the bottom to the top. 
\label{moszczenica}
}
\end{figure}

Measurements of the quasi-free $pn\to pn\eta$ reaction have been performed 
at the cooler synchrotron COSY~\cite{cosy} at the Research Center J\"ulich in Germany
by means of the COSY-11 detector setup~\cite{cosy11} 
presented schematically  in Fig.~\ref{moszczenica}.
The proton beam with a nominal momentum of $p_{beam}~=~2.075$~GeV/c,
(corresponding to the kinetic energy of $T~=~1.339$~GeV) 
was circulating through the  
deuteron cluster jet target~\cite{dombrowski} with an
 areal density of about 10$^{14}$ atoms/cm$^2$.
Due to the relatively weak binding energy of the deuteron ($\sim$~2.2~MeV) which is by more than two orders 
of magnitude smaller than the kinetic energy of the bombarding protons and due to 
the relatively large average distance between the proton and the neutron ($\sim$3~fm~\cite{garcon}),
it is quite probable that the beam proton interacts with the nucleon from the deuteron as if it would be a free
particle. 
Such quasi-free proton-neutron interactions were tagged by the registration of the low energy proton 
moving upstream the beam. In the analysis
the proton emerging from the deuteron is considered 
as a spectator which leaves the interaction region with the Fermi momentum 
undisturbed by the final state interaction with the other reaction products.
In this approach we assume that the measured spectator proton 
was on its mass shell already at the collision moment
and that the matrix element 
for the production of the $\eta$ meson by the  beam proton off the neutron  bound in the deuteron
is identical to that for the free $pn\to pn\eta$ reaction. 
This assumption is supported by a theoretical investigation~\cite{kaptari}
and was already confirmed by various experiments~\cite{calen2,tofpn1,tofpn2,jaegle,triumf}.
In particular 
it was proven that even in the case of the pion production it is valid up to a Fermi momentum
of 150 MeV/c~\cite{tofpn1}.

The Fermi motion of nucleons causes the smearing of the 
total 
center-of-mass energy for the proton-neutron reaction ($\sqrt{s_{pn}}$).
The resultant distribution of the excess energy for the quasi-free 
$pn\to pn\eta$ reaction (Q~$\equiv$~$\sqrt{s_{pn}}-m_{p}-m_n-m_{\eta}$) 
is broader than  50~MeV~\cite{hab,c11pn,czyzyk}, and 
therefore to achieve an accuracy of Q in
the order of a few MeV it is important to reconstruct the four-momentum vector
of the interacting neutron on the event-by-event basis.
Such an accuracy is mandatory for  near threshold studies
where the cross sections may vary by more than an order 
of magnitude within 
an excess energy range of a few tens of MeV~\cite{review}.
Therefore, the spectator proton was registered not only to tag the 
quasi free proton-neutron reaction but also 
to determine
the four-momentum vector of the reacting neutron.
The usage  of the thin cluster target (10$^{14}$ atoms/cm$^2$) 
and mounting of the spectator detector
inside the beam pipe ensured the negligible distortion of the momentum of 
the measured spectators.  
Based on the aforementioned spectator model and 
on the energy and momentum conservation, 
the four momentum of the target neutron
$\mathbb{P}_{t}^{n} \equiv (E^{n}_{t}, \vec{p}^{\, n}_{t})$ 
at the moment of the collision was derived 
from the measured spectator 
four momentum $\mathbb{P}_{sp}\equiv (E_{sp}, \vec{p}_{sp})$ according to the following formulae:
$E^{n}_{t} = m_d - E_{sp}$, \ and \ 
$\vec{p}^{\, n}_{t} = -\vec{p}_{sp}$, 
\label{sekowa}
where $m_d$ denotes the mass of the deuteron.  
The absolute momentum of the beam proton was determined based on the known beam optics and the
frequency measurement of the circulating beam~\cite{cosy}.
The outgoing proton from the $pn\to pn X $ reaction
 was separated from the beam in the magnetic field
of the dipole and it was 
identified by the independent measurement of its momentum and velocity
using drift chambers and scintillators.                   
The momentum of outgoing proton was reconstructed with a 
precision of 6~MeV/c (standard deviation)~\cite{c11PhysRev}
by tracing  its trajectory 
in the magnetic field of the dipole back to the interaction point.
The outgoing neutron was measured by means of the neutral particle detector,
which delivered information about the position and time at which the registered neutron induced a hadronic reaction.
Finally, the $\eta$ meson was 
identified via the missing mass technique,
where the four-momentum of an unobserved object produced in the quasi-free $pn \to pnX$ reaction
was determined as:
$  \mathbb{P}_{X} = \mathbb{P}_{t}^{n} + \mathbb{P}_{beam} - \mathbb{P}_{proton} - \mathbb{P}_{neutron}$.
The evaluation of the four-momentum vectors
of the outgoing neutron and the spectator proton as well as
the functioning of detectors used for their registration will be described
more detailed in the following  sections. 
Whereas for the detailed description of 
other detectors used in this experiment 
the reader is referred to the previous publications of the COSY-11 group~\cite{cosy11}.

\subsection{Neutron detector}
The neutral particle detector 
is positioned at a distance of 7.36~m
from the interaction region. It consists of 24 modules, each  built out of
11 plastic scintillator plates 
interlayed by 11  plates of lead
each with dimensions of 270~$\times$~90~$\times$~4~mm$^3$~\cite{czyzyk,joanna}. 
Each detection 
unit is read out on its upper and lower edge by photomulitipliers.
The modules are arranged into five layers as it is schematically shown in Fig.~\ref{moszczenica}.
In front of the first layer, in order to permit a separation between charged and neutral particles,  
an additional scintillator counter (often referred to as a veto detector)
built out of four overlapping modules with  dimensions of 400~$\times$~200~$\times$~4~mm$^3$
has been positioned~\cite{rozek}.
\begin{figure}[h]
\begin{center}
  \includegraphics[width=4.6cm]{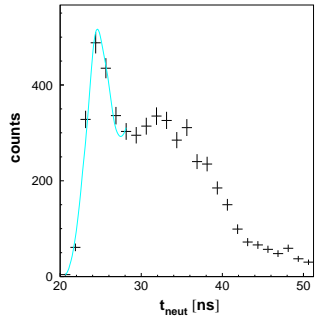}
\end{center}
\caption{
Distribution of 
the time-of-flight measured
between the target and the neutral particle detector. 
}
\label{neutron}
\end{figure}
The determination of the four momentum of the neutron is based on the 
time of flight between the interaction point and the hit position 
in the neutral particle detector~\cite{czyzyk,joanna,rozek}.
The time of the reaction is deduced from the measurements of the proton velocity, 
its trajectory and the time when it crossed the S1 counter.
The obtained distribution 
of the time of flight 
is presented in Fig.~\ref{neutron}. 
A clear peak around  24.5~ns is associated to the 
$\gamma$ quanta originating from decays 
of e.g. $\pi^0$ mesons
which are produced copiously 
in the proton-deuteron
reactions.
For the $pn\to pn\eta$ reaction, at a beam momentum of $p_{beam}~=~2.075$~GeV/c, 
the time of flight expected for neutrons is larger than 32~ns.  
The identification of neutron allowed to determine
its energy and momentum vector based on the time-of-flight and the hit position, the latter was 
defined as the center of that module whose signal was produced as a first one.

\subsection{Spectator detector}
For the determination of the kinetic energy and scattering angle of
the spectator protons a dedicated silicon detector system 
has been used~\cite{bilger}. 
It consists of four double-layered modules each with a thickness of 300~$\mu$m.
The front layer (the one closer to the beam) 
contains 18 silicon strips each with an active
area of 20$\times$5~mm$^2$, while the back layer contains 6 silicon strips 
with active area of 20$\times$18~mm$^2$. 
A schematic view of the spectator detector arrangement is presented in Fig.~\ref{moszczenica}.
The modules are positioned 
at a distance of about 5~cm from the interaction region  
and their active area covers about 22\% 
of the full solid angle. The arrangement was a compromise between the maximum coverage
of the solid angle, angular resolution and the technical needs for the 
installation~\cite{c11pn,czyzyk}.
The detector  
has been 
used previously for experiments at the CELSIUS storage ring by the PROMICE/WASA
collaboration~\cite{bilger}.
\begin{figure}[h]
\begin{center}
  \includegraphics[width=3.8cm]{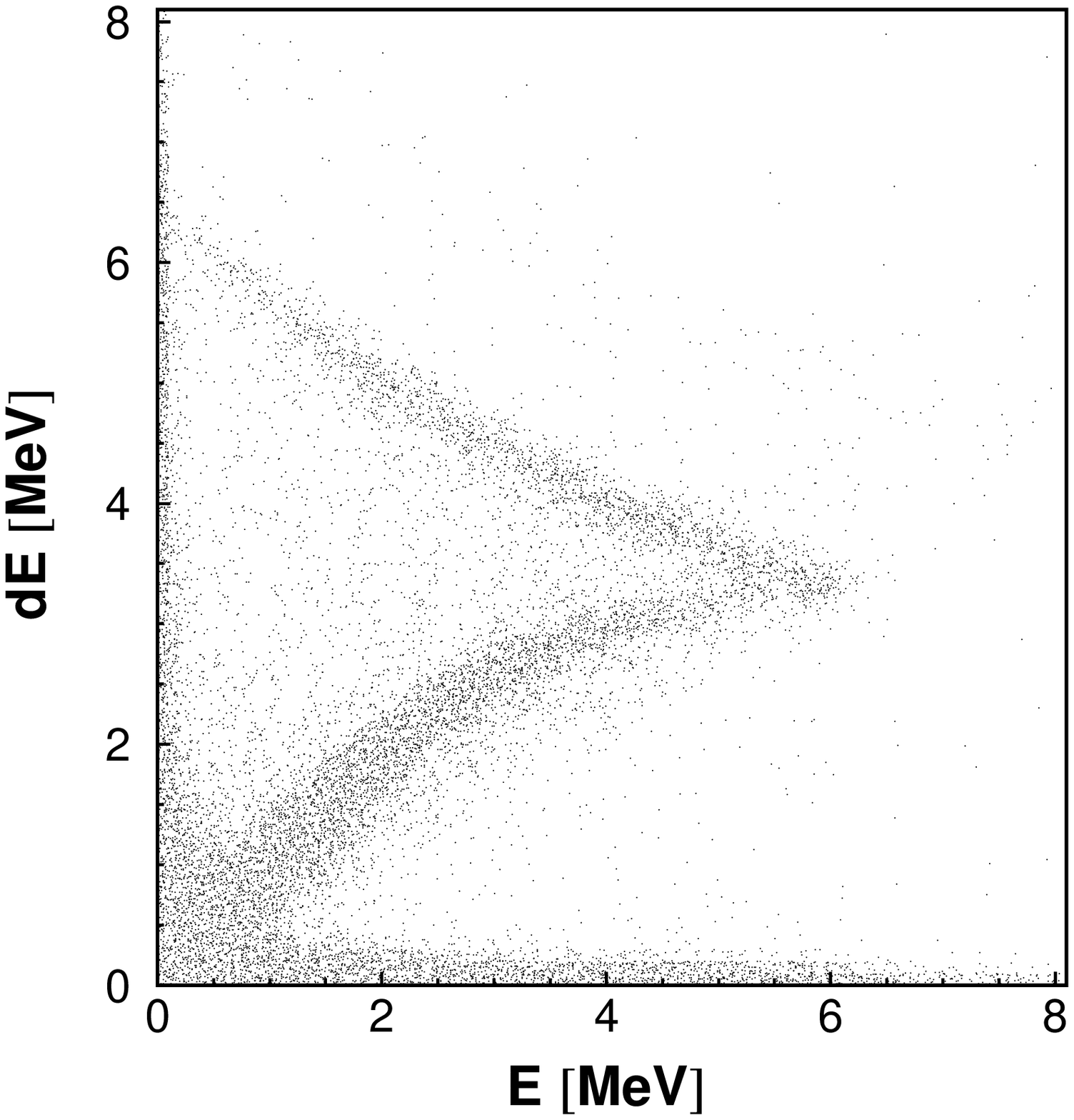}
  \hspace{-0.4cm}
  \includegraphics[width=3.8cm]{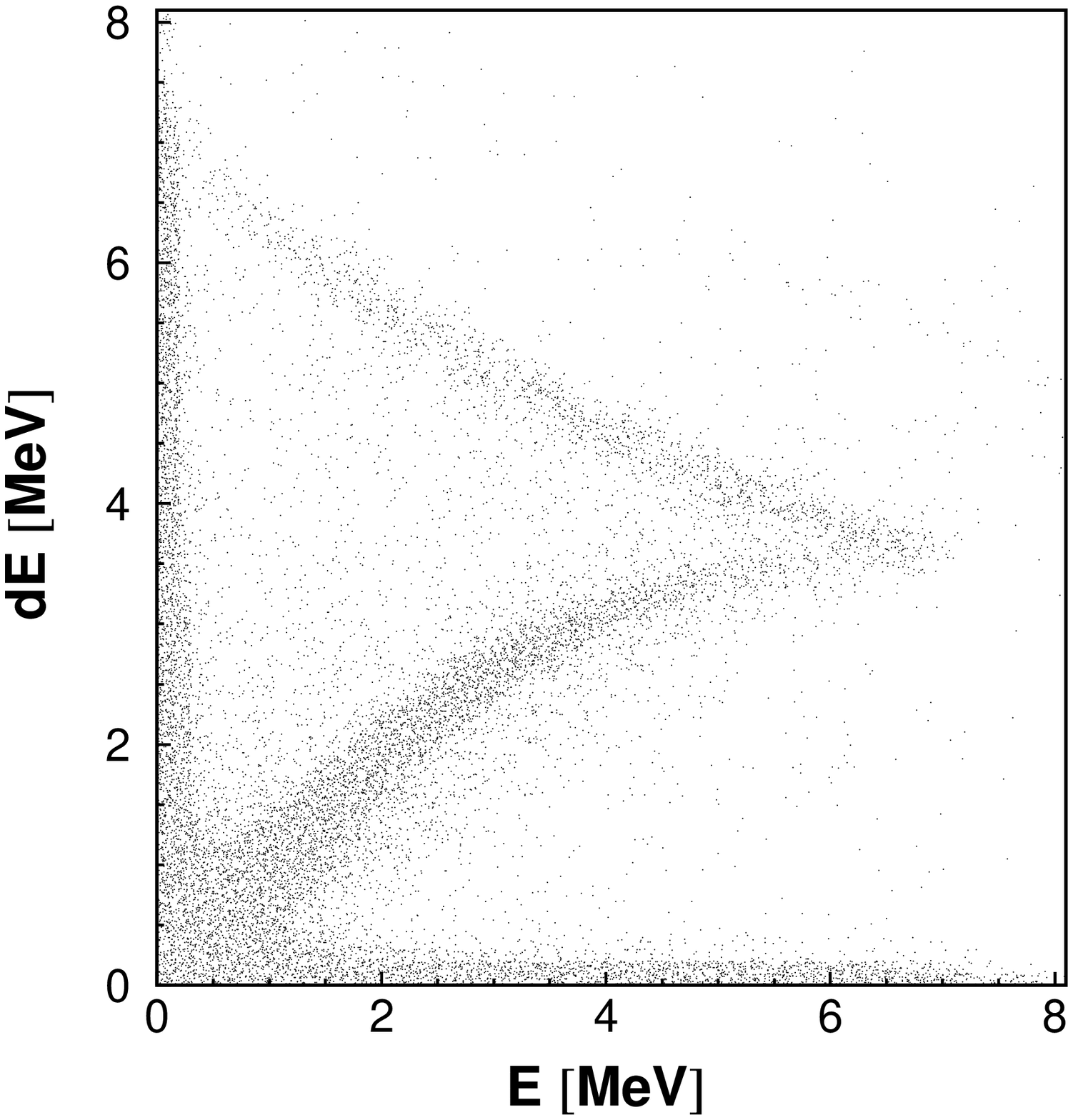}
\end{center}
\caption{dE--E plot for events registered in the spectator silicon detector
for two 
pairs placed under different angles with respect to the interaction region. 
Signals from deuterons are not observed
since deuterons cannot be emitted
towards the detector (upwards to the proton beam) 
due to the kinematics.
\label{banan}
}
\end{figure}
Figure~\ref{banan} presents  observed spectra of the energy 
losses deposited  in the front layer of the silicon 
detector (dE axis) versus the energy deposition in the back layer (E axis)
for two pairs (out of 72 pairs) positioned at  different distances from the interaction region. 
Both distributions show clear bands from the low energy protons.
Differences in shapes of the proton bands in the left and right panel
of Fig.~\ref{banan}
are due to the fact that the effective thickness of the 
silicon pads as seen by the particles outgoing from the interaction region
varies as a function of the particle's emission angle.
These variations allowed to check the position and orientation of the
spectator detector relative to the interaction region. 
For this aim we have 
compared the experimental dE-E distributions 
to the spectra simulated for different 
positions of the specator detector.
By examining the 
$\chi^2$ distribution we have determined the arrangement of 
modules  with an accuracy of $\pm$~1~mm, consistent with the
nominal values based on the geometrical design of the setup.
The accuracy of $\pm$~1~mm is fully sufficient in view of the
size of the  stream of the deuteron target with a diameter 
of 9~mm~\cite{dombrowski}.

\subsection{Determination of the excess energy and the production yields}
\begin{figure}[h]
\begin{center}
  \includegraphics[width=4.5cm]{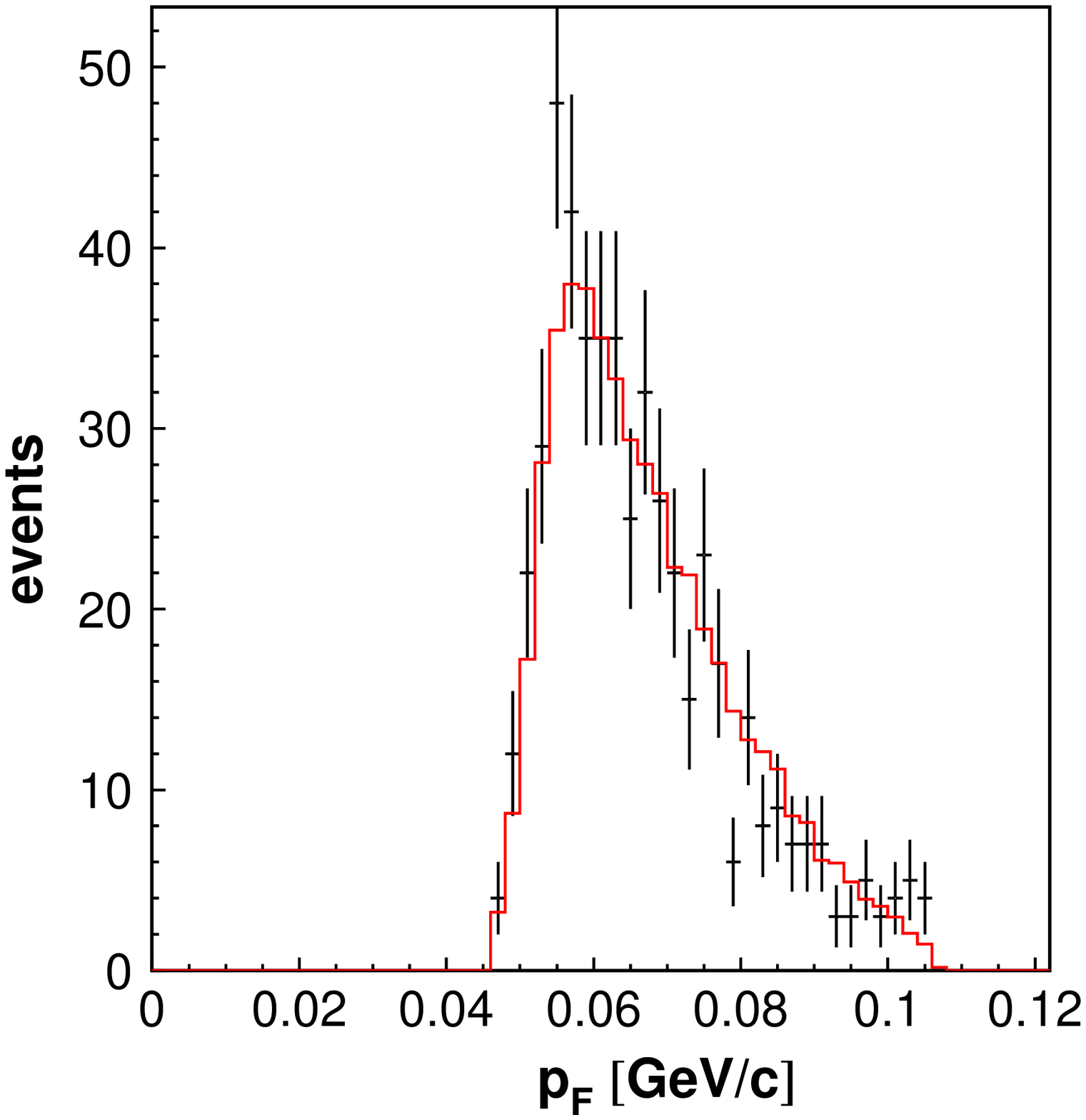}
  \includegraphics[width=4.5cm]{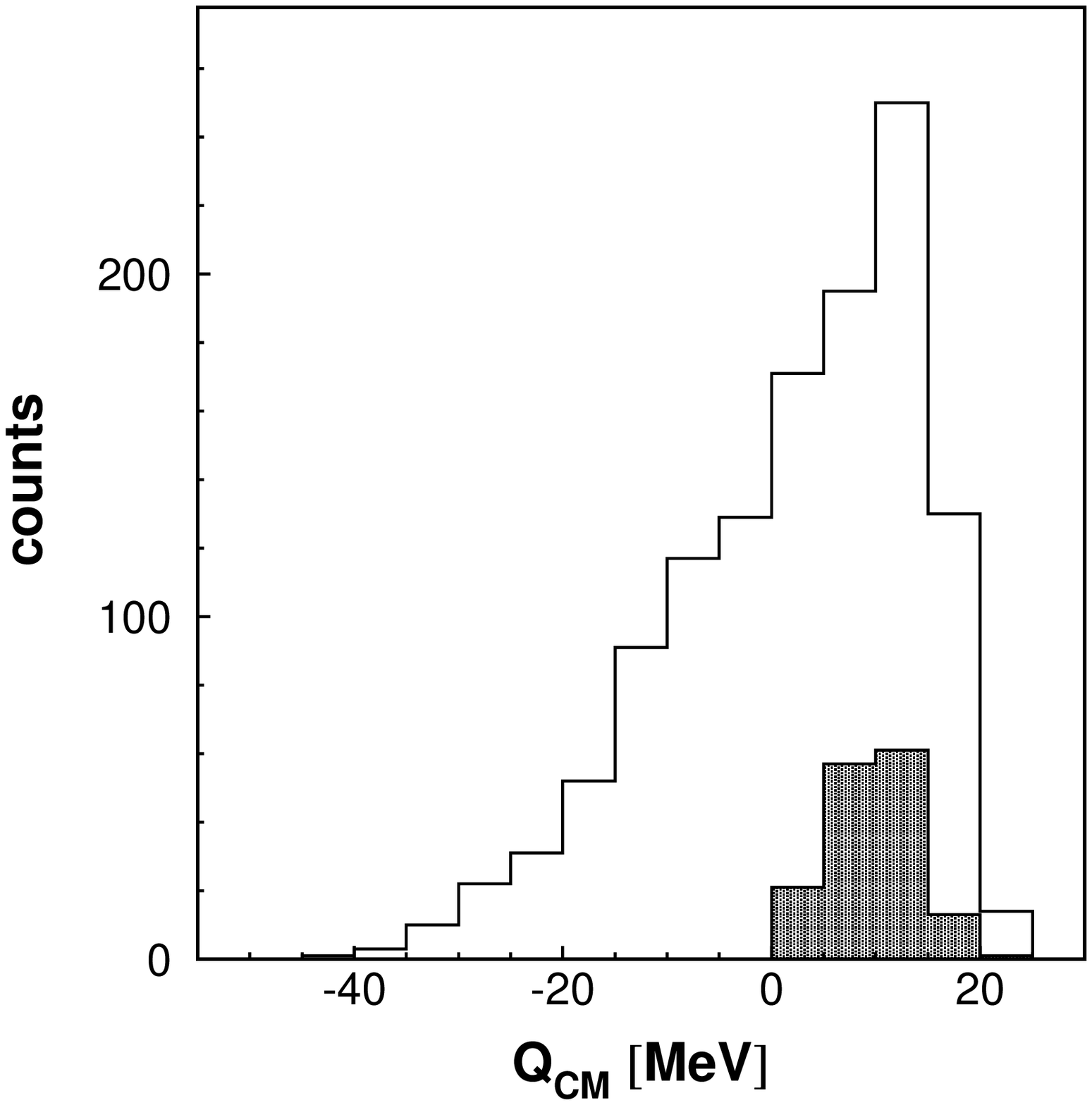}
\end{center}
\caption{
(left)  Points with error bars denote
experimental distribution of the momentum of the spectator protons.
Full line represents the Monte-Carlo spectrum of the spectator momentum
resulting from  the 
Fermi momentum 
distribution~\cite{paris} of nucleons inside the deuteron 
convoluted with the acceptance and efficiency of the COSY-11 detector setup.
(right) 
Experimental distribution of the excess energy Q with respect to the 
$pn\eta$ system  determined for the 
$pn\to pnX$ reaction (solid line) and for the 
$pn\to pn\eta$ events (shaded histogram).
}
\label{lipinki}
\end{figure}

A spectrum of the momenta 
of 
particles identified as spectator protons,
as measured in this experiment, 
is shown in the left panel of Fig.~\ref{lipinki}.
In addition to the data points, this figure shows a superimposed histogram
depicting expectations derived assuming that the momentum distribution
of nucleons inside the deuteron is given by the Paris nucleon-nucleon potential~\cite{paris}.
A good agreement between the experimental and simulated spectra raised the confidence
to the method used.
In the experiment we were able to register the spectator protons with momenta 
ranging from 50~MeV/c to circa 115~MeV/c. 
The energy losses 
for protons with momenta lower than 50~MeV/c  are not separable from the noise range
and the spectators with momenta higher than 115~MeV
were not taken into account since in the analysis we used the back
layer of the spectator detector as a veto for the reduction of signals from 
charged pions.
The noise cut has been performed for each silicon pad separately, 
analysing the spectra of energy loss 
triggered by a pulser with  
a frequency of 1~Hz
used concurrently with other experimental triggers.

The measurement of the spectator momentum and as a consequence the determination of the 
total energy for the quasi-free proton-neutron collisions allowed to establish 
the  excess energy Q with respect to the $pn\to pn\eta$ process for each event.
The determined experimental distribution is shown in the right panel of Fig.~\ref{lipinki}.
Its shape results from i) the genuine excitation function for the $\eta$ meson
and the multi-meson production in the proton-neutron collision, ii) from the 
Fermi momentum distributions of the nucleons in the deuteron target, and iii) from the acceptance
and efficiency of the COSY-11 apparatus.
  
For negative excess energies Q the $\eta$ meson cannot be created
and hence all events for Q~$<$~0 originate from the quasi-free pions production.
For positive Q additionally to the production of pions also the 
$pn\to pn\eta$ reaction can occur. This, however cannot be identified 
on the event by event basis since the COSY-11 detector system does not allow
for the efficient registration of the decay products of the produced mesons.
Therefore,  in order to disentangle  between  pions and the $\eta$ meson production 
we have
grouped  the collected data according to the excess energy
and for each sample the number of $pn\to pn\eta$ reaction was extracted
based on the missing mass distribution. 
Optimizing the statistics and resolution, the range of excess energy above the 
$\eta$ meson production threshold has been divided into 
intervals of $\Delta$Q~=~5~MeV width.  The width of the bin corresponds
to the accuracy (FWHM)  of the determination of the  excess energy
which was estimated based on Monte-Carlo simulations
taking into account the size of the target, 
the spread of the beam momentum 
as well as the horizontal ($\sigma$(x)~$\approx$~2~mm) 
and vertical~($\sigma$(y)~$\approx$~4~mm) beam size~\cite{pawel-nim}.
As a result for the $pn\to pn\eta$ reaction  near the threshold
we derived a standard deviation for the excess energy  of $\sigma$(Q)~=~2.2~MeV 
which is to be compared with 1.8~MeV 
obtained at similar conditions at the PROMICE/WASA setup~\cite{bilger}.

For Q~$>$~0 missing mass spectra
have been established for each interval of $\Delta$Q separately. 
Next, 
a corresponding background distribution 
has been constructed from all events with Q~$<$~0~MeV,
following the procedure 
described in details in 
a dedicated article~\cite{pawel7}. 

Subsequently, 
to  each 
distribution for Q~$>$~0 
the corresponding background spectrum has been normalized 
for mass values less than 0.25 GeV/c$^2$.
In this missing mass region events correspond to a single
pion production for which a production cross section stays nearly constant 
when the excess energy changes by a few tens of MeV since with respect to the single pion 
the energy is high above the threshold.
In order to gain more confidence to the procedure
we have also performed an alternative normalisation, requiring that after the background
subtraction the ratio of the integral of missing mass experimental 
spectrum between 0.3 and 0.5~MeV/c$^2$ to the integral between 0.5~MeV/c$^2$ and
the kinematical limit is the same as the corresponding ratio of integrals
obtained from Monte-Carlo simulations.
Both normalisation methods resulted in a statistically consistent result.
\begin{figure}[h]
\begin{center}
  \includegraphics[width=4.8cm]{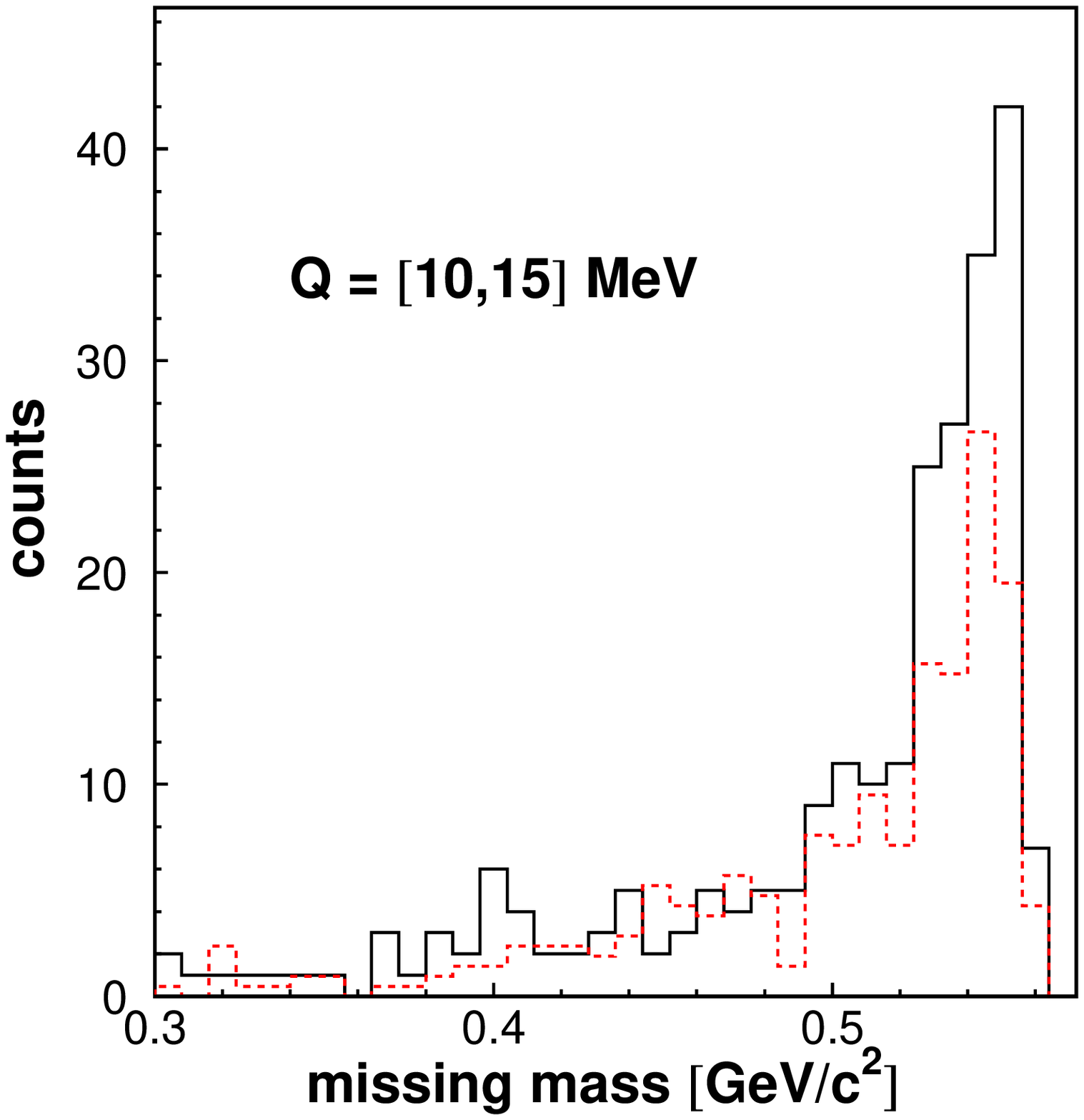} \hspace{-0.4cm}
  \includegraphics[width=4.8cm]{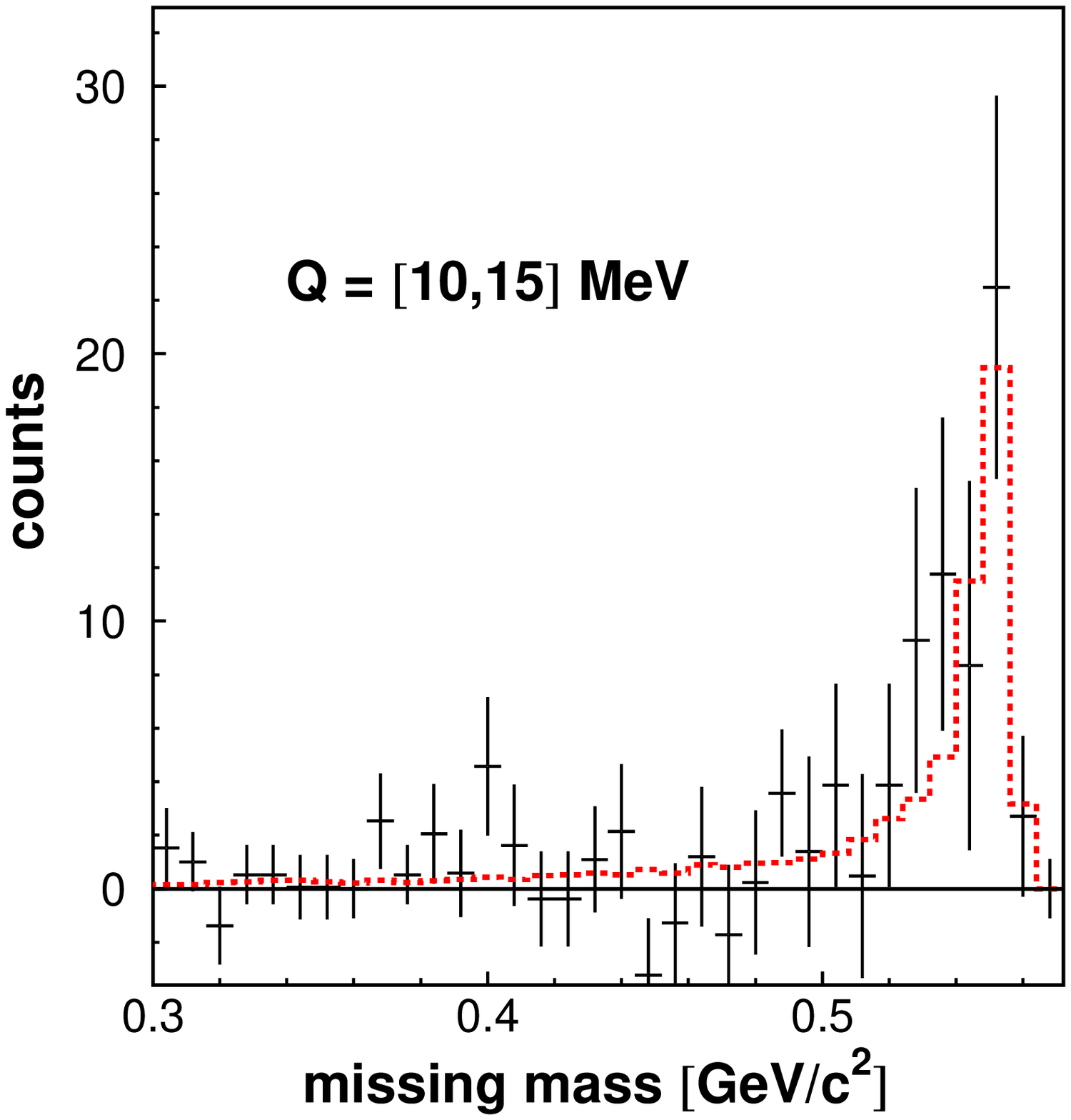}
\end{center}
\caption{
(left) The solid line denotes the missing mass distribution of the $pn\to pnX$ reaction
       determined for the excess energy range from 10 to 15~MeV, and the dashed line shows
       the background distribution constructed from events with Q~$<$~0
       according to the method described in reference~\cite{pawel7}.
(right) Missing mass distribution of the $pn\to pn\eta$  reaction determined 
        by subtraction of histograms from the left panel.
        Vertical bars indicate statistical errors only. 
        Superimposed dashed line denotes result of simulations 
        normalized to the data in amplitude.
\label{gorlice}
}
\end{figure}
As an example, the left panel of Fig.~\ref{gorlice}
shows  the experimental 
missing mass spectra 
for
signal~(solid line)
and background~(dashed line) 
corresponding to the excess energy interval
between 10 and 15~MeV. 
The resulting background subtracted distribution
for the quasi-free $pn\to pn\eta$ reaction
is presented in the right panel of Fig.~\ref{gorlice}.
The number of $pn\to pn\eta$ events has been extracted from this spectrum by 
fitting to it the 
simulated distribution with the amplitude 
as the only free parameter.
The simulation 
was based on the GEANT-3 code~\cite{geant} containing 
the exact geometry
of the COSY-11 detector system as well as the precise map of the 
magnetic field of the dipole magnet. Also the momentum and spatial beam 
spreads, multiple scattering of particles, and other known physical and instrumental 
effects have been taken into account~\cite{hab,cosy11,pawel-nim}. 
The calculations take into
account also the proton-neutron 
final state interaction~\cite{czyzyk} which changes the acceptance by 5\%.
Finally, in order to obtain 
the missing mass spectrum (dashed line in Fig.~\ref{gorlice}~(right))
a simulated detector response was analyzed in the same way 
as the experimental data. 

\subsection{Luminosity}
The reference reaction used for the purpose of calculating the
luminosity was the quasi-free elastic scattering of protons.
The measurement was based on the registration of the momentum vector
of the forward scattered proton and the direction of motion of the recoil proton.
The recoil protons were registered in the scintillator counter S4 
and the granulated 
silicon detector Si$_{mon}$, while the forward scattered protons 
were registered by the stack of drift chambers D1 and D2, and the scintillator
array S1~(see Fig.~\ref{moszczenica}). For triggering 
the coincidence between signals from S1 and S4 scintillators was required. 
Events corresponding to the elastically scattered protons have been identified
on the basis of the distributions 
of the transversal versus the parallel momentum components 
of the forward scattered proton, on which a 
signal from the $pp\to pp$ reaction  appears as a clear 
enhancement of the density of events around the expected
kinematical ellipse~\cite{pawel-nim,aiplumi}.
In order to determine the integrated luminosity,
numbers of elastically scattered protons 
registered in 
modules of 
the S1 detector 
were compared to the inner products
of the differential cross sections 
and the probability
density of the distribution of the Fermi momentum computed
for 
the solid angles covered by the relevant S1 detection units.
For more detailed description 
the interested reader is referred to the dedicated article~\cite{aiplumi}.
The values of integrated luminosity determined in  
four independent 
detection modules of the S1 detector 
are  statistically consistent with the 
average amounting to: 
L~=~(208~$\pm$~3)~$\cdot$~nb$^{-1}$, where 
the quoted error stands for the statistical uncertainty only.

\section{Results}
\begin{figure}[h]
\begin{center}
  \includegraphics[width=6cm]{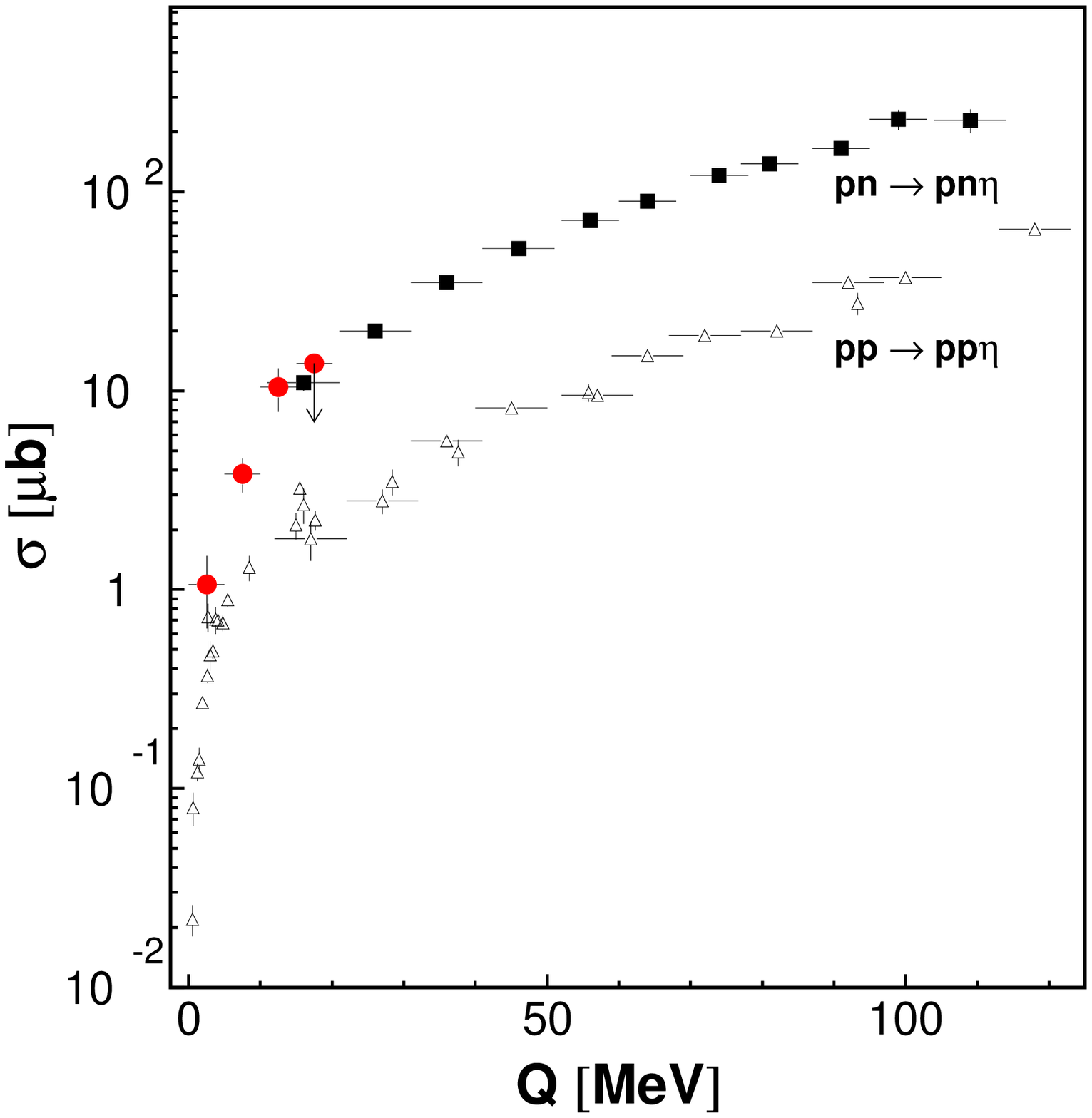}
  \includegraphics[width=6cm]{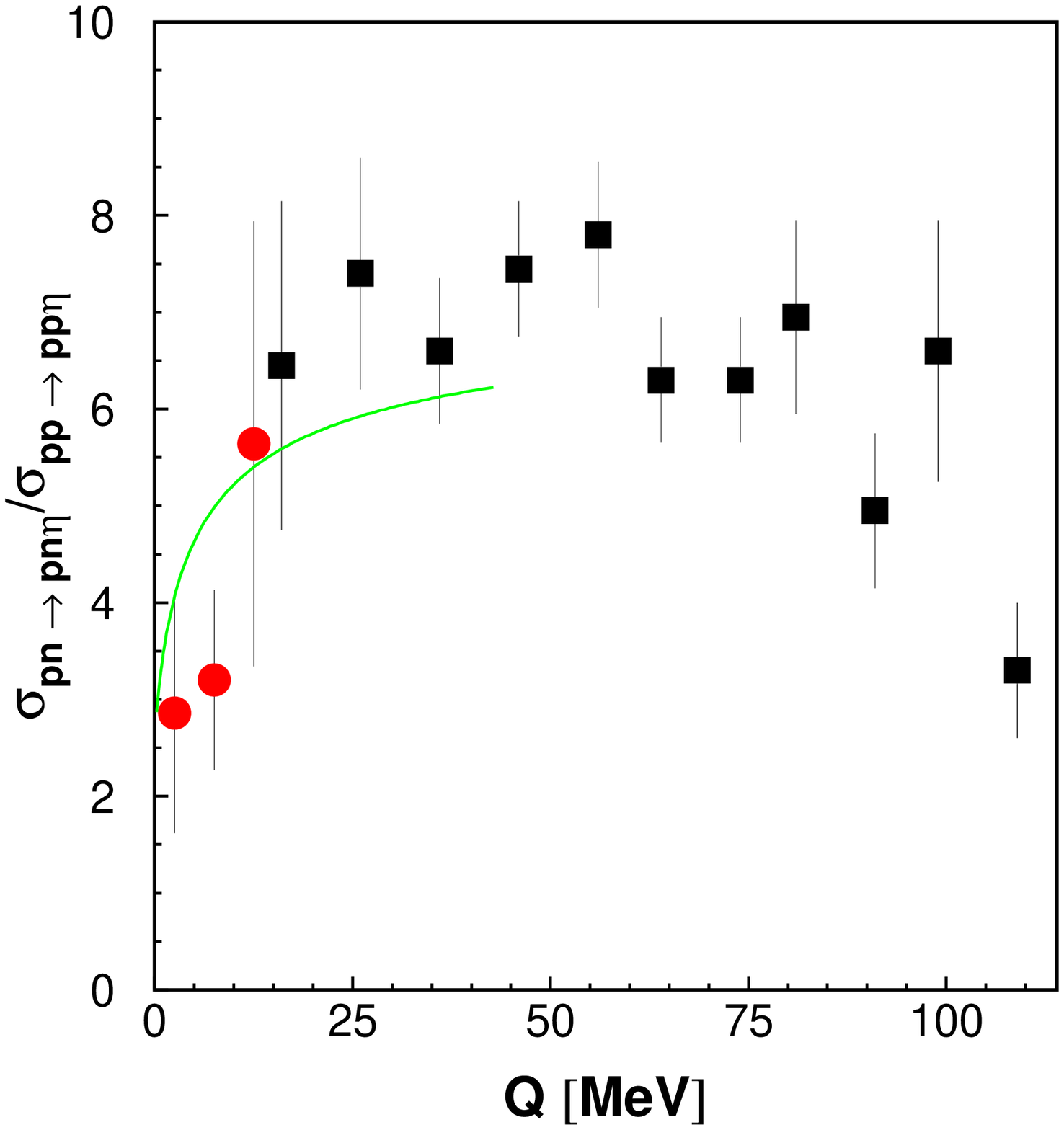}
\end{center}
\caption{(Left)
Excitation functions for the 
near threshold
$\eta$ meson production via reactions $pn\to pn\eta$ 
(filled symbols)
and $pp\to pp\eta$ 
(open symbols)~\cite{smyrski,chiavassa,calen1,calen2,c11PhysRev,hibou}.
Results of  this article, presented as full circles, 
are consistent with 
values of cross sections determined 
by the CELSIUS/WASA group (filled squares)~\cite{calenpn}.
(Right)
Ratio of the total cross sections for the $pn\to pn\eta$
and $pp\to pp\eta$ reactions.
A superimposed line indicates a result of the fit taking into account 
the final state interaction of nucleons.
\label{kryg}
}
\end{figure}

Total cross sections determined for the quasi-free $pn\to pn\eta$ reaction
are given in Table~\ref{magura} and are presented 
in the left panel of
Fig.~\ref{kryg}. 
\begin{table}[h]
\caption{ Total cross sections for the $pn\to pn\eta$ reaction as a function of the excess energy.
Given are the statistical and systematic uncertainties, respectively.        
The energy intervals correspond to the binning applied.
                \label{magura}}
\begin{center}
\begin{tabular}{|cc|}
\hline
Q [MeV]  &  $\sigma$ [$\mu$b] \\
\hline
\hline
(0,5]    &  1.06~$\pm$~0.42~$\pm$~0.21 \\
(5,10]   &  3.81~$\pm$~0.74~$\pm$~0.76 \\
(10,15]  &  10.4~$\pm$~2.6~$\pm$~2.1 \\
(15,20]  &  upper limit of 13.7 at 90\% CL\\
\hline
\end{tabular}
\end{center}
\end{table}    
Vertical error bars shown in Fig.~\ref{kryg} denote the 
statistical uncertainty only, while the horizontal bars represent the 
size of the excess energy bins for which the total cross section 
values were extracted. 
In addition the total systematical uncertainty was estimated to be about 20\%.
It originates from ten independent contributions which where added
in quadrature, and which 
were estimated by 
varying (in the analysis of the data and simulations of the acceptances)
the values of parameters used for the description of the experiment.
The studies revealed that:
i) changing the  global time offset of the neutron detector,  within one standard deviation 
of its statistical uncertainty, results in a change of the cross sections
by $\pm$5\%, ii) the variation of the cut on the noises in the spectator detector 
within the limits of  energy resolution resulted in an error of $\pm$7\%,
iii) changing the position of the spectator within the range 
of expected uncertainty ($\pm$1~mm)
resulted in an error of $\pm$15\%, iv) varying in simulations the
beam momentum resolution arbitrarily by 
$\pm$1~MeV/c around its nominal value of 3.5~MeV/c 
resulted in changes of cross sections values by $\pm$3\%, 
v) the changes in simulations of the time resolution 
of the neutron detector (0.4~ns)
by $\pm$0.2~ns gave $\pm$5\% error in the cross section. 
Additionally we have also taken into 
account an uncertainty due to the method used 
for the background subtraction amounting to~3\%~\cite{pawel7},
and systematic errors of the luminosity determination
discussed and estimated in details in reference~\cite{aiplumi},
where it was established that the significant contributions
originate from the method of the background subtraction (3\%),
and from the overall normalisation error of the reference data~(4\%)~\cite{edda}. 
We took also into account  uncertainty~of~$\pm$2\%~\cite{czyzyk} connected to 
the calculations of the Fermi momentum distribution.
This uncertainty was established as the  difference between results determined
using the Paris~\cite{paris} and the CDBONN~\cite{cdbonn} potentials.
The other important effect considered 
is the reabsorption
of the produced meson by the spectator proton which 
is proportional 
to the average of the inverse square of the distance between two nucleons 
in the deuteron and which  
reduces the cross section by a factor of about 3\%~\cite{chiavassa2}.
Therefore we increased the determined cross sections by a factor 
of 1.03 assuming conservatively the uncertainty of this correction
to be not larger than $\pm$1\% of the cross section.
Another nuclear effect 
decreasing the total 
cross section by about 4.5\%~\cite{chiavassa}, 
referred to as the ``shadow effect'',
was not taken into account because
the reduction of the beam flux  seen by the neutron, 
due to the shielding by a spectator proton, 
is expected to be the same for the quasi-elastic scattering 
which was used for the determination of the luminosity. 

Finally the uncertainty of the detection 
efficiency, 
predominantly due to the uncertainty in the efficiency of the neutron
registration, is estimated to be not larger than~$\pm$5\%. 
The computation of the efficiency and acceptance of the COSY-11 detection system 
is based on the GEANT-3~\cite{geant} simulation packages which uses the 
GEANT-FLUKA subroutines for the calculations of the hadronic interactions. 
For the beam momentum of 2.075~GeV/c and the conditions of the detection with the COSY-11 setup
the kinetic energy of neutrons from the $pd\to p_{sp}pn\eta$ reaction  ranges from ~200 to 430~MeV.
The resulting efficiency of the 44~cm thick COSY-11 neutral particle detector varies in this neutron
energy  range from 51\% to 56\%~\cite{rozek}. This is in a very good agreement with expectations
based on the efficiencies determined for LAND~\cite{land} and other calorimeters~\cite{rozek,calorimeters}.

The right panel in Fig.~\ref{kryg} shows the ratio of the total cross sections
for the $pn\to pn\eta$ and $pp\to pp\eta$ reactions plotted as a
function of the excess energy. 
It can be  seen 
that this ratio 
falls down at 
lower values of~Q. 
To large extent, this behavior may  plausibly be explained 
by the difference in strength of the proton-proton 
and proton-neutron FSI~\cite{colinpriv2008}.
An influence of the nucleon-nucleon interaction on the 
shape of the excitation function
for the $pn\to pn\eta$ and $pp\to pp\eta$
reactions 
may be well described by the closed analytical formula derived 
by F{\"a}ldt and Wilkin~\cite{fw1,fw2} which implies that:
\begin{equation}
\frac{\sigma(pn\to pn\eta)}{\sigma(pp\to pp\eta)} = 0.5 + C (\frac{\sqrt{\epsilon_{pp}}+\sqrt{\epsilon_{pp}+Q}}{\sqrt{\epsilon_{pn}}+\sqrt{\epsilon_{pn}+Q}})^2, 
\label{krakow2}
\end{equation}
where 
$\epsilon_{pn}~=~2.2$~MeV and $\epsilon_{pp}~=~0.68$~MeV~\cite{pawel2} are the corresponding
``binding'' energies of the pn bound and pp virtual
states, respectively~\cite{fw1}. 
We have fitted the function
given by equation~\ref{krakow2} (with C as the only free parameter) 
to the data in the excess energy range 
from 0 to 40~MeV 
where the higher partial waves of the proton-proton
system are suppressed~\cite{hab}. 
The result 
is presented in Fig.~\ref{kryg} as the solid line and  explained to some extent
the observed decrease of the ratio at threshold.

\section{Summary}
Using the
COSY-11 detector setup we have conducted
measurements of the total cross sections for the quasi-free
$pn\to pn\eta$ reaction in the very close-to-threshold region.
The experiment has been performed investigating the $\eta$ meson 
production on a neutron bound in a deuteron target. 
The quasi-free proton-neutron reactions were tagged by the registration
of the spectator proton.
The Fermi 
momentum distribution of the nucleons inside the deuteron 
has been accounted for in the calculations of the integrated luminosity
as well as in the determination of the total cross section.
The derived total cross sections are consistent 
in the overlapping excess energy range 
with the previous measurement
performed by the CELSIUS/WASA group. 
At the threshold the determined total cross section for the 
$pn\to pn\eta$ process exceeds
the total cross section for the $pp\to pp\eta$ reaction  by a factor 
of three in contrast to the  factor of six observed  
for higher excess energies. The observed decrease 
may be assigned to some extent to the different
energy dependence of the proton-proton and proton-neutron
final state interactions~\cite{colinpriv2008}. 
A slight bump-like structure in the ratio of the $pn\to pn\eta$
to $pp\to pp\eta$ cross sections with a flat maximum at an excess 
energy of about 50~MeV could be due to the resonance N$^*(1535)$ 
(m(N$^*$)~-~m$_\eta$~-~m$_{nucleon}$~$\approx$~49~MeV)
indicating that coupling of this resonance 
to the neutron--$\eta$ can be stronger than to the proton-$\eta$ channel.
\vspace{1pc}

\section{Acknowledgements} 
We are grateful to Colin Wilkin for many useful discussions
and for the interpretation of the decrease of the cross sections ratio.
We acknowledge the great 
help when installing the 
neutron and spectator detectors 
received from
E. Bia{\l}kowski, 
O. Felden, 
G. Friori,
W. Migda{\l},
and D. Proti{\'c}. 
The work was partially supported by the
European Community-Research Infrastructure Activity
under the FP6 programme (Hadron Physics,
RII3-CT-2004-506078), by
the Polish Ministry of Science and Higher Education under grants
No. 3240/H03/2006/31  and 1202/DFG/2007/03,
and by the German Research Foundation (DFG).

\end{document}